\documentclass{aastex62}
\usepackage{color}
\shortauthors{Chung et al.}

\begin{document}
\title{Formation of Blue-cored Dwarf Early-type Galaxies in a Cluster Environment: a Kinematical Perspective}


\author{Jiwon Chung}
\affil{Department of Astronomy and Space Science, Chungnam National University, Daejeon 34134, Korea; jjiwon1114@gmail.com, screy@cnu.ac.kr}
\affil{Research Institute of Natural Sciences, Chungnam National University, Daejeon 34134, Korea}

\author{Soo-Chang Rey}
\affil{Department of Astronomy and Space Science, Chungnam National University, Daejeon 34134, Korea; jjiwon1114@gmail.com, screy@cnu.ac.kr}

\author{Eon-Chang Sung}
\affil{Korea Astronomy \& Space Science Institute, Daejeon 34055, Korea}

\author{Suk Kim}
\affil{Department of Astronomy and Space Science, Chungnam National University, Daejeon 34134, Korea; jjiwon1114@gmail.com, screy@cnu.ac.kr}
\affil{Korea Astronomy \& Space Science Institute, Daejeon 34055, Korea}
\affil{Center for Galaxy Evolution Research, Yonsei University, Seoul 03722, Korea}

\author{Youngdae Lee}
\affil{Korea Astronomy \& Space Science Institute, Daejeon 34055, Korea}

\author{Woong Lee}
\affil{Department of Astronomy and Space Science, Chungnam National University, Daejeon 34134, Korea; jjiwon1114@gmail.com, screy@cnu.ac.kr}




\begin{abstract}

The presence of blue cores in some dwarf early-type galaxies (dEs) in galaxy clusters suggests the scenario of late-type galaxy infall and subsequent transformation into red, quiescent dEs. We present Gemini Multi-Object Spectrographs long-slit spectroscopy of two dEs with blue cores (dE(bc)s), EVCC 591 and EVCC 516, located at the core and outskirt of the Virgo cluster, respectively. We obtained their internal kinematics along the major axis out to, at least, $\sim$ 1 effective radius. EVCC 591 shows evidence of a kinematically decoupled core (KDC) with a size of 2$\arcsec$ (160 pc), exhibiting an inverted pattern for velocity with respect to the main body of its host galaxy. The rotation curve of the stellar component in the inner region of EVCC 591 is steeper than that in the rest of the galaxy. On the other hand, overall velocity profiles of stellar and ionized gas components of EVCC 516 show no signature of significant rotation. The occurrence of a KDC and zero rotation in the internal kinematics along with the central star formation support the scenario of gas-rich dwarf-dwarf mergers in the formation of these two dE(bc)s. Furthermore, evolution of dE(bc)s in a cluster environment into ordinary dEs with KDCs is possible based on their structural properties. We suggest that at least some of the dE(bc)s in the Virgo cluster were formed through dwarf-dwarf mergers in lower density environments before they subsequently fell into the cluster; they were then quenched by subsequent effects within the cluster environment.

\end{abstract}


\section{Introduction}

Dwarf galaxies, the most numerous low-luminosity systems in the universe \citep{fb94}, have been considered as the building blocks of massive galaxies in $\Lambda$ cold dark matter hierarchical merging scenarios \citep[e.g.,][]{WR78,WF91}. In particular, dwarf early-type galaxies (dEs), including dwarf ellipticals and dwarf lenticulars, are the numerically dominant galaxy class especially in clusters and groups, outnumbering any other galaxy type, and are very rare in an isolated environment \citep[e.g.,][]{fb94,sandage_B_T85,Binggeli+88}. The dEs play an important role in understanding galaxy evolution in the cluster as their shallow gravitational potential well makes them more susceptible to external influences compared to more massive galaxies. Thus, they are ideal laboratories to test various mechanisms that govern galaxy evolution in a cluster environment. \\

Despite their rather unspectacular, simple appearance, a surprising complexity and diversity in terms of their characteristics in a cluster (e.g., structure, kinematics, stellar content, and spatial distribution) has become evident \citep[e.g.,][]{Jerjen+00,lisker06a,lisker06b,lisker07,Michielsen+08,Chilingarian09,Toloba+09,toloba+11,toloba14a,toloba14b,toloba15,paudel+10,janz+12,janz+14}. For example, a systematic study of Virgo dEs revealed several subclasses of dEs showing different underlying substructures \citep{lisker06a,lisker06b,lisker07}. The varying range of properties strongly suggests that the majority, if not all, of dEs are most likely descendants transformed from late-type progenitors owing to external perturbations in the cluster environment (see \citealt{Lisker09} and references therein). \\

In this respect, transitional dwarf galaxies--those on the way to becoming red dEs from late-type progenitors--are the most attractive targets to investigate the effect of the environment on the evolution of galaxies. One of the most interesting transitional objects is the dEs showing blue cores at their centers \citep{DeRijcke+03,Cellone&Buzzoni05,gu06,lisker06a,Tully&Trentham08,DeRijcke+13,pak+14}. These blue-cored dEs, dubbed as dE(bc)s, are classified as dE based on their morphology; however, they exhibit blue central colors along with central irregularities and dust structures caused by recent or ongoing star formations \citep[e.g.,][]{deLooze+10}. The colors of the outer parts of dE(bc)s are comparable to the integrated colors of normal, red dEs without blue cores  \citep{lisker06a,pak+14}. Therefore, dE(bc)s have a significant positive color gradient in the color profiles. Most dE(bc)s in the Virgo cluster exhibit color differences of more than 0.2 mag between the central and outer regions \citep{lisker06a}. In general, dE(bc)s in the Virgo cluster span a range of $r$-band absolute magnitude (-18 $<$ M$_r$ $<$ -14)  and $g-i$ color ( 0.6 $<$ $g-i$ $<$ 1) \citep{meyer14}. The dE(bc)s reach a fraction of more than 15$\%$ of the bright (m$_B$ $<$ 16 mag) dE population in the Virgo cluster \citep{lisker06a}. Recent studies also revealed that dE(bc)s are more ubiquitous in a group environment \citep[e.g.,][]{Tully&Trentham08,pak+14}.  \\

A few different formation mechanisms of dE(bc)s in a cluster have been proposed. The two most discussed ones are ram-pressure stripping and galaxy harassment in a cluster \citep{moore98,Mastro+05,Boselli08a,boselli08b,lisker06a,Lisker09}. Ram-pressure stripping for infalling late-type star-forming galaxies by the hot intracluster medium (ICM) within the cluster could be responsible for the formation of dE(bc)s, in which the interstellar gas of the galaxy is rapidly removed, but some of it can be retained in the central region \citep{Boselli08a}. Galaxy harassment also appears to transform the infalling late-type galaxies into galaxies with early-type morphology \citep{moore98}. In this case, it is possible that dE(bc)s could be formed via star formation in the central regions of late-type galaxies triggered by gravitational interaction with other galaxies. This process gives rise to excess of gas density in the center by funneling gas into the central region  \citep{moore98}. Another plausible scenario for the formation of dE(bc)s is the merger between dwarf galaxies with available gas reservoirs. Dwarf-dwarf merging can trigger central bursts of star formation via the infall of gas, which forms massive compact cores dominated by young stellar population. During this process, the galaxy appears as a blue compact dwarf (BCD) galaxy and subsequently become a galaxy morphologically similar to a dE(bc) after the last BCD phase \citep[e.g.,][]{Bekki08,Watts&Bekki16}. However, all these proposed processes explain only part of the observed properties of dE(bc)s in the cluster \citep{lisker06a}. \\

On the other hand, the internal kinematics of dwarf galaxies provides the most crucial information for discriminating different formation and evolution mechanisms, as stellar kinematics keeps memory of the processes that occurred in the course of galaxy evolution. The angular momentum of a galaxy is conserved in a ram-pressure stripping event in the cluster environment, and a rotationally supported system has been observed \citep{Boselli08a,boselli08b,Benson+15}. However, in the multiple gravitational interactions with other cluster galaxies and with the potential of the cluster, the galaxy is rapidly heated. This leads to an increase in the velocity dispersion of the galaxy and a decrease in its rotation \citep{Mastro+05, Benson+15, toloba15}. Moreover, the possible existence of kinematically decoupled cores (KDCs) in dEs supports the scenario of mergers in galaxy pairs \citep{toloba14a}. A distorted rotation curve is also considered as a signature of strong tidal interactions \citep[e.g., dE satellites of M31,][]{Geha06, geha+10}. Recent observational studies revealed that dEs in the cluster have a large variety of kinematic features including non-rotators, slow/fast rotators, and anomalous rotation curves \citep{Rys+13, toloba14b, toloba15}. \\

The kinematical properties of dE(bc)s in a galaxy cluster will also contribute to the better understanding of the formation mechanism of dE(bc)s, primarily in a dense environment. However, mainly because of the faint luminosity (M$_r$ $>$ -18 mag) and low surface brightness ($\mu$$_r$ $\geq$ 21 mag arcsec$^{-2}$ at effective radius) of dE(bc)s \citep{meyer14}, a systematic study on the internal kinematics for extensive samples of dE(bc)s has not been conducted. As part of our study in response to this situation, we carried out long-slit spectroscopic observations of two dE(bc)s in the Virgo cluster using a Gemini 8-m telescope. This paper is organized as follows. In Section 2, we describe the two sample dE(bc)s, spectroscopic observations, and data reduction procedures. In Section 3, we present the results of the stellar and ionized gas kinematics of galaxies and also discuss their possible formation scenario and star formation history. We summarize our results and conclusions in Section 4. Throughout this paper, we assume a distance of 16.5 Mpc to the Virgo cluster \citep{Jerjen+04,mei07}.  \\

\section{DATA AND ANALYSIS}
\subsection{Target Galaxies}

The dEs in the Virgo cluster were chosen from the Extended Virgo Cluster Catalog \citep[EVCC;][]{kim14}, covering an area of the Virgo cluster larger than that of the classical Virgo Cluster Catalog \citep[VCC;][]{binggeli85}. As part of our study on dE(bc)s in the Virgo cluster using the EVCC, the dE(bc)s were classified as dEs with a clear positive color gradient based on the radial  $g$-$i$ color profile of galaxies obtained from images of the Sloan Digital Sky Survey (SDSS) data release 12 \citep[DR12;][]{Alam+15}. We selected two target dE(bc) galaxies, EVCC 591 (VCC 870) and EVCC 516, which are classified as nucleated dwarf lenticular (dS0, N) and nucleated dwarf elliptical (dE, N) galaxies, respectively, in the EVCC. EVCC 591 was also defined as the ``main sample" of dE(bc) by \citet{lisker06a}. These two galaxies have relatively high surface brightness ($\mu$$_r$  $\sim$ 22 mag arcsec$^{-2}$) at their effective radii, which help obtain reliable signal-to-noise (S/N) ratios of their spectra. \\

The SDSS $g$, $r$, and $i$ composite color images of two galaxies show bluer colors in their central regions with respect to the remaining parts of the galaxies (top panels of Figure 1). This is in accordance with the significant positive color gradient in the  $g$-$i$ color profiles, indicating the existence of young stellar populations and distinct star formation history in the galaxy center (middle panels of Figure 1). The $g$-$i$ color differences between the inner and outer regions are approximately 0.4 mag and 0.7 mag for EVCC 591 and EVCC 516, respectively. In the case of EVCC 591, the region with the most blue $g$-$i$ color shows an offset of $\sim$ 1$\arcsec$ from the photometric center of the galaxy owing to the off-centered nucleus of this galaxy. The SDSS spectra of galaxies clearly exhibit ongoing star formation activity at their centers within a 3$\arcsec$ fiber diameter, which corresponds to a physical size of 240 pc at a distance of the Virgo cluster. The spectrum of EVCC 591 displays star formation activity with strong H$\alpha$ and H$\beta$ emission lines superposed on a blue continuum, and EVCC 516 shows more active star formation with prominent, additional emission lines (e.g., [O III]$\lambda$5007) (see bottom panels of Figure 1). The overall strengths of the emission lines of EVCC 516 are larger than those of EVCC 591. \\

We derived the integrated ultraviolet (UV) star formation rate (SFR) using near-ultraviolet (NUV; 1750-2800 \AA) flux from the Galaxy Evolution Explorer GR6 data \citep{martin+05}. The UV SFRs of two target galaxies are computed using the relation given by \citet{kennicutt98}. The SFRs(NUV) of EVCC 591 and EVCC 516 are $\sim$ 0.013 M$_\odot$ yr$^{-1}$ and $\sim$ 0.020 M$_\odot$ yr$^{-1}$, respectively. Further, we estimated the stellar mass of each galaxy from its $i$-band magnitude and $g-i$ color using the relation of \citet{bell03}; $\sim$ 10$^{9.1}$ M$_\odot$ and $\sim$ 10$^{8.8}$ M$_\odot$ for EVCC 591 and EVCC 516, respectively. Finally, NUV specific SFRs, which are SFRs normalized by stellar masses, of EVCC 591 and EVCC 516 are $\sim$ 1.03 $\times$ 10$^{-11}$ yr$^{-1}$ and $\sim$ 3.17 $\times$ 10$^{-11}$ yr$^{-1}$, respectively, indicating that the star formation of EVCC 516 is $\sim$ 3 times more active than EVCC 591. \\

The two target galaxies are defined as certain members in the EVCC located inside a spherical symmetric infall model in a plot of radial velocity versus clustercentric distance of galaxies (see \citealt{kim14} for details). They are located in different environments within the Virgo cluster. EVCC 591 is located at the cluster core, but EVCC 516 is on the outskirts of the Virgo cluster, outside the X-ray distribution from the ROSAT (red contours in Figure 2) and the region of the VCC (dot-dashed contour in Figure 2). EVCC 591 and EVCC 516 are located at projected distances of 0.37 Mpc ($\sim$ 0.24 R$_{virial}$) and 2.58 Mpc ($\sim$ 1.67 R$_{virial}$) from the Virgo cluster center, respectively, considering the distance of the Virgo cluster and virial radius (R$_{virial}$) of Virgo A \citep[1.55 Mpc,][]{Ferra12}. The basic characteristics of the target galaxies are summarized in the Table 1. \\

\subsection{ Surface Photometry}

We performed surface photometry of the galaxies using the background-subtracted optical images in the $g$, $r$, and $i$ bands from the SDSS DR12. Contaminating foreground and background objects were masked manually. We derived the surface brightness profiles of the galaxies using the IRAF $\it{ELLIPSE}$ task \citep{Jedrzejewski87}. We fixed the center of the isophote to the photometric center of the galaxy adopted from the EVCC, while the position angle and ellipticity were set as free parameters in all bands. \\

In Figures 3a and c (from top to bottom), we present $r$-band profiles of the surface brightness, position angle, ellipticity, and $a$$_4$/$a$ parameter returned from the ellipse fitting. The observed surface brightness profiles (gray solid lines) of galaxies were best-fitted using a combination (red dashed line) of the Gauss function for the central nucleus (blue dashed line) and S\'ersic function for the overall galaxy structure with $n$=1.6 for EVCC 591 and $n$=1.4 for EVCC 516 (orange dashed line). It is evident that the observed surface brightness profiles show the departures from a single S\'ersic fit at the central region of the galaxy, while the S\'ersic fit for the overall galaxy is consistent with the typical dE profile. Note that it is common to fit the surface brightness profiles of dEs with one or more additional components in addition to a single S\'ersic profile (see \citealt{janz+12,janz+14} for details). We also estimated effective radius R$_{eff}$ (12$\arcsec$.3 $\pm$ 0$\arcsec$.1 for EVCC 591 and 10$\arcsec$.5 $\pm$ 0$\arcsec$.1 for EVCC 516), mean effective surface brightness $<$$\mu$$>$$_{eff,r}$  (21.70 $\pm$ 0.13 mag arcsec$^{-2}$ for EVCC 591 and 21.59 $\pm$ 0.12 mag arcsec$^{-2}$ for EVCC 516), and uncertainty-weighted mean ellipticity  (0.262 $\pm$ 0.003 for EVCC 591 and 0.274 $\pm$ 0.004 for EVCC 516) within the R$_{eff}$ based on the $r$-band surface brightness profile excluding data inside the seeing radius ($\sim$ 0$\arcsec$.7 based on the median value of point spread function width in SDSS $r$-band\footnote{\url{http://www.sdss.org/dr12/imaging/other\_info/}}). The $a$$_4$/$a$ values of EVCC 591 and EVCC 516 are $\sim$ 0 at all radii, while slightly disky isophotes (i.e., $a$$_4$/$a$ $>$ 0) are observed in their inner regions. \\

To concatenate the kinematical properties of galaxies with possible photometric substructures, we created residual images of galaxies by subtracting two-dimensional smooth model images constructed by ellipse fitting from the original SDSS $r$-band images. In Figures 3b and d, we also illustrate residual images of EVCC 591 and EVCC 516, respectively. A notable feature in the residual image of EVCC 591 is that the nucleus, shown as a bright blob within the long slit, is off-centered (by $\sim$1$\arcsec$) toward the southwest from the galaxy center (red cross). \citet{Binggeli00} also noted that the nucleus of this galaxy shows a clear displacement of 1$\arcsec$ from the galaxy center within the lens-like structure. From the residual image of EVCC 591 obtained using a Next Generation Virgo Cluster Survey deep image with a much higher spatial resolution, we also confirmed the features of the off-centered nucleus (not shown). In the case of EVCC 516, the nucleus coincides well with the photometric center. The residual image reveals a number of knots around the galaxy center and one knot at r $\sim$ -4$\arcsec$ is located in the long-slit region of our observations.\\

\subsection{Observations and  Data Reduction}

The spectroscopic observations were carried out using the Gemini Multi-Object Spectrographs \citep[GMOS;][]{Hook04} mounted on the 8.1-m Gemini north telescope in the long-slit mode, during the 2015A semester (Program ID: GN-2015A-Q-202). The slit was oriented along the major axis of the galaxy and its width was set to 1$\arcsec$. The seeing varied between 0$\arcsec$.5 and 0$\arcsec$.6 during the observations. The total exposure time was 10,800 s for each galaxy, which was split into nine exposures of 1200 s. The GMOS detector comprises three 2048 $\times$ 4608 EEV CCDs with a pixel size of 13.5 $\times$ 13.5 $\mu$m$^2$. We applied 2 $\times$ 2 CCD binning, yielding a spatial resolution of 0.1454 arcsec pixel$^{-1}$ and reaching an S/N ratio of 15 or better at the galaxy center. We used the B1200 G5301 grism covering the wavelength range 4680 - 6150 \AA\ with a reciprocal instrumental resolution R=3744 (i.e., dispersion of 0.23 \AA\  pixel$^{-1}$). The spectral resolution obtained from the full width at half-maximum of the copper argon (CuAr) emission line is 1.68 \AA, which corresponds to the velocity resolution of 97 km s$^{-1}$. Two template stars (HD 92049 (K0 III) and HD 55068 (G0 Ib)) and one standard star (EG 131 (DBQ5)) were observed in the same spectroscopic setup as the galaxies. Bias frames, dome flat fields, and copper argon (CuAr) arcs were also taken for calibrations. In Table 2, we summarize the instrumental configuration used for the observations. \\

Data reduction was performed following the standard procedure for long-slit spectroscopy using the IRAF Gemini/GMOS package version 1.13. (1) Initial reduction of all CCD frames including overscan/bias subtraction and flat-field correction was performed through the $\it{GSREDUCE}$ task. (2) Nine 1200 s exposure frames for each galaxy were combined using the $\it{GEMCOMBINE}$ task. (3) We used the $\it{GSWAVELENGTH}$ task to perform wavelength calibration from the CuAr arc frames. Then the wavelength calibration was applied to the object frames using the $\it{GSTRANSFORM}$ task. (4) The galaxy frames were sky-subtracted using the $\it{BACKGROUND}$ task, which uses background samples from off-object areas. (5) Finally, one-dimensional spectra along the spatial direction were extracted from the two-dimensional spectra using the $\it{SCOMBINE}$ task. The spatial width (i.e., the number of CCD rows binned) for each extracted one-dimensional spectrum was increased with the radius from the galaxy center to achieve a minimum S/N ratio per \AA\ of $\sim$5 for all galactocentric radii in the spectral region of Mg$b$. This allowed us to obtain reliable profiles for the velocity and velocity dispersion along the major axis out to, at least, $\sim$ 1 R$_{eff}$.\\

We obtained the radial velocity and velocity dispersion at each galactocentric radius by obtaining the cross-correlation function along the major axis of the galaxies using the $\it{FXCOR}$ task in IRAF. Inadvertently, a part of the Mg$b$ absorption lines of the two template stars and a standard star were not observed as these lines were located in the CCD gap. The absence of Mg$b$ absorption lines of the template stars can degrade the accuracy of the cross correlation. However, the Mg$b$ absorption lines were observed at 5194 \AA\ and 5190 \AA\ for EVCC 591 and EVCC 516, respectively, avoiding the CCD gap as the two galaxies have relatively high radial velocities (1194 km s$^{-1}$ and 955 km s$^{-1}$ for EVCC 591 and EVCC 516, respectively). Therefore, to obtain the velocity profile of the stellar population in a galaxy with respect to the galaxy center, the galaxy spectrum at each galactocentric radius was cross-correlated with the spectrum of the galaxy center as a template. On the other hand, the velocity dispersion of the stellar population was estimated using two template stars. The IRAF $\it{SPLOT}$ task was used for measuring the radial velocity of ionized gas in the galaxies by Gaussian fitting to the [O III]$\lambda$5007 emission line. As our two dE(bc)s have H$\beta$ emission lines falling in the H$\beta$ Balmer absorption line, we could not obtain the age information of the stellar population in these galaxies. Furthermore, owing to the absence of Mg$b$ absorption lines of a standard star, information on the metallicity of galaxies could not be obtained as well. \\

\section{RESULTS AND DISCUSSION}
\subsection{Kinematics}

We explored the internal kinematics of stellar and ionized gas components extracted from the Mg$b$ absorption and [O III]$\lambda$5007 emission lines, respectively, for the two target galaxies. In Figure 4, we present kinematical and other observable profiles of EVCC 591 (left panels) and EVCC 516 (right panels) as a function of the distance from the center of the galaxy along the major axis; i.e., residual images, the two-dimensional spectrum around [O III]$\lambda$5007 emission line, velocity profiles of stellar and ionized components with respect to the systemic velocities of galaxies, and stellar velocity dispersion profiles (from top to bottom). EVCC 591 exhibits a substantial rotation feature in the stellar velocity profile (red filled circles in Figure 4c). The maximum rotation velocity (V$_{rot}$) at R$_{eff}$ is approximately 19.25 km s$^{-1}$, which was determined as the mean of the representative values around R$_{eff}$ along opposite semimajor axes. The velocity profile of the ionized gas component (green open circles in Figure 4c) is only shown for limited locations along the major axis due to the absence or very weak emission lines of [O III]$\lambda$5007 in other regions. The velocity dispersion profile is rather flat up to R$_{eff}$, with a mean value of 44.01 km s$^{-1}$ within the effective radius.\\

We determined V$_{rot}$/$\sigma$$_{eff}$ ($\sim$ 0.44) of EVCC 591, where V$_{rot}$ is the velocity at R$_{eff}$ and $\sigma$$_{eff}$ is the mean value of the velocity dispersion within R$_{eff}$, and then the integrated two-dimensional (V/$\sigma$)$_{eff}$ ($\sim$ 0.17) within R$_{eff}$ using the equation (V/$\sigma$)$_{eff}$ = (0.39 $\pm$ 0.02)V$_{rot}$/$\sigma$$_{eff}$ proposed by \citet{toloba15}. \citet{toloba15} found a correlation between the internal kinematics of dEs and their location in the Virgo cluster; according to this correlation, at a given ellipticity, the dEs in the inner regions of the cluster rotate slower (i.e., smaller (V/$\sigma$)$_{eff}$) than their counterparts in the outer regions (see Figure 3 in \citealt{toloba15}). In the plane of (V/$\sigma$)$_{eff}$ vs. $\epsilon$$_{eff}$ , where $\epsilon$$_{eff}$ is the average of ellipticities within R$_{eff}$, EVCC 591 is located slightly above the demarcation line between the fast and slow rotating systems defined by \citet[][see also dashed line of Figure 3 of \citealt{toloba15}]{Emsellem+07,Emsellem+11}  at a given $\epsilon$$_{eff}$. \\

One interesting feature in the velocity profile of EVCC 591 is that a kinematic disturbance is shown at $\sim$ -1$\arcsec$ from the galaxy center, where an off-centered nucleus is observed. The nucleus exhibits an inverted velocity pattern (at approximately 5 km s$^{-1}$) with respect to the main body of the galaxy. This KDC has a size of $\sim$2$\arcsec$ ($\sim$ 160 pc). The velocity dispersion of the kinematically decoupled nucleus does not deviate from the rest of the galaxy. On the other hand, the rotation curve of the stellar component in the inner region (-5$\arcsec$ $<$ R $<$ +2$\arcsec$) of EVCC 591 appears to be steeper than the outer region of the galaxy. The slopes of the rotation curves of the inner (-5$\arcsec$ $<$ R $<$ +2$\arcsec$) and outer (R $<$ -7$\arcsec$ and R $>$ +5$\arcsec$) regions are 4.6 km s$^{-1}$ arcsec$^{-1}$ and 1.6 km s$^{-1}$ arcsec$^{-1}$, respectively. This suggests that EVCC 591 contains a rapidly rotating central stellar component with a concentration of mass embedded in the main body of the galaxy, which shows relatively slow or negligible rotation at larger radii (see Sec. 3.2.1 for details).  \\

Contrary to the case of EVCC 591, the overall velocity profiles of the stellar and ionized gas components of EVCC 516 show no signature of significant rotation (Figure 4g). However, the ionized gas appears to exhibit a sign of rotation with a rotation velocity of $\sim$ 10 km s$^{-1}$  in the central region. A disturbance is shown at R $\sim$ -4$\arcsec$ in the velocity profiles of the stellar and ionized gas components, which coincides with an identified knot in the residual image (see Figure 4e). This knot has a velocity difference of $\sim$ 20 km s$^{-1}$ from the center of the galaxy. Another disturbance is also shown at R $\sim$ +4$\arcsec$ in the velocity profile of the ionized gas, while no such a feature was observed in the stellar component. No particular structure was identified at this location of the long-slit region, while one knot is located just below this position in the residual image. The velocity dispersion profile shows a flat distribution, with a mean value of 47.38 km s$^{-1}$ within the effective radius. \\

\subsection{Formation Scenario and Star Formation History}

It is well predicted that the merging of two gas-rich dwarfs can trigger central starbursts of BCDs by gas inflow and then form massive young, compact cores embedded in old stars \citep[e.g.,][]{Bekki08}. After the last BCD phase, the morphology of the merger remnant with a positive age gradient in the numerical simulation of \citet{Bekki08} is comparable to those of the observed dE(bc)s \citep[e.g.,][]{lisker06a,pak+14}. In this regard, while a few different formation channels of dE(bc)s in the cluster have been proposed, we discuss galaxy merging as a relevant formation process of our two target galaxies in this paper. \\

\subsubsection{EVCC 591}

One of the interesting features in the velocity profile of EVCC 591 is the presence of a KDC in its off-centered nucleus. While the KDC was studied in numerous bright, massive early-type galaxies \citep{bacon01,cappell11,sanchez12}, the existence of this structure was reported for only a handful of dEs \citep{thomas06, Chilingarian+07, koleva+11, toloba14a,Guerou15}. In terms of the suggested processes for transformation into dEs from late-type galaxies in the cluster, environmental effects within the cluster could be related to the KDC formation in dEs. As a dominant environmental effect in the central region of the cluster, ram-pressure stripping has direct effect on the gas in the progenitor galaxy. However, this process did not obviously affect its stellar kinematics and is therefore not relevant to the formation of a KDC. In the case of galaxy harassment, strong tidal encounters affect the stellar kinematics of the outer envelope of the galaxy as shown in N-body simulations of the encounter between a dwarf galaxy and a massive elliptical galaxy \citep[e.g.,][]{Gonzalez-Garcia+05}. However, harassment can mostly produce large-scale counter-rotating signatures at larger radii ($>$ 1.5 R$_{eff}$) of the dwarf galaxy, rather than a small-scale ($\ll$ R$_{eff}$) structure of the KDC observed in dEs. Moreover, these kinematical structures are formed under limited conditions in which only $\sim$ 1$\%$ of the dwarf galaxies in a cluster like the Virgo cluster may show counter-rotating features resulting from harassment (see \citealt{Gonzalez-Garcia+05} for details). \\

The merger between galaxies has been widely suggested as the appropriate mechanism for KDC formation in early-type galaxies \citep{tsatsi15, krajnovic+15, Schulze17}. If a galaxy has its orbit in a direction opposite to that of the other one in the merging event, this could explain the kinematically decoupled feature in the velocity profile \citep{Chilingarian+07}. Large, kpc-scale KDCs in large fractions of slow-rotating massive early-type galaxies show similar old ($>$ 8 Gyr) ages with respect to the host galaxy suggesting a dry merger for KDC formation \citep{McDermid06}. On the other hand, the majority of compact ($<$ a few hundred pc) KDCs with a very young ($<$ 5 Gyr) stellar population are found in fast-rotating massive early-type galaxies \citep{McDermid06}. Two dEs (VCC 1183 and VCC 1453) in the Virgo cluster are also fast-rotating systems and their compact ($<$ 500 pc) KDCs show a younger and possibly more metal-rich stellar population than the main body of the galaxy; this suggests that KDCs are likely formed from the merger between gas-rich progenitors or in manner similar to centrally localized star formation (see \citealt{toloba14a} for details). It is worth noting that EVCC 591 appears to share similar properties with VCC 1183 and VCC 1453 in the sense that the KDC of EVCC 591 is very compact ($\sim$ 160 pc) and has a bluer color (i.e., younger age) than the outer part of the galaxy. \\

In addition to the presence of the KDC, another interesting kinematical feature of EVCC 591 is the steep central rotation curve compared to the rest of the galaxy. It is known that BCDs also exhibit central rotation curves significantly steeper than those in the outer regions, implying a strong central concentration of the dynamical mass \citep{meurer+98,lelli+12a,lelli+12b,lelli+14a,lelli+14b,koleva14}. The merger between gas-rich dwarf irregular galaxies has been suggested as one of the most supporting mechanisms for the central concentration of mass in the BCDs, as large amounts of gas are funneled into the central region of the merger remnant (\citealt{Bekki08,bekki15,Watts&Bekki16} and references therein). Recently, in their numerical simulations on dwarf-dwarf mergers, \citet{Watts&Bekki16} confirmed that the BCDs with central gas concentration formed through major merging can effectively produce steep rotation curves and high surface brightnesses in their central regions, which is consistent with previous observational results \citep{lelli+14a}. \\

In this regard, in the plane of the inner circular-velocity gradient (V$_{R_d}$/R$_d$) against the central surface brightness ($\mu$$_0$), we compared EVCC 591 with various types of dwarf galaxies extracted from \citet{lelli+14a} (see Figure 5). V$_{R_d}$/R$_d$ provides a direct estimate of the dynamical mass density in a galaxy center, where R$_d$ is the exponential scale length of a stellar body and V$_{R_d}$ is the circular velocity at R$_d$ (see \citealt{lelli+14a} for details). Following the procedures of \citet{lelli+14a}, we estimated the values of V$_{R_d}$/R$_d$ (= 55.95 $\pm$ 5.57 km s$^{-1}$ kpc$^{-1}$) and $\mu$$_0$ (= 21.18 $\pm$ 0.12 mag arcsec$^{-2}$) for EVCC 591. All values were calculated from the $R$-band luminosity profile of EVCC 591 constructed using the SDSS $g$- and $r$-band profiles and conversion equation of \citet{windhorst+91}. Figure 5 shows a strong correlation between V$_{R_d}$/R$_d$ and $\mu$$_0$ of the sample galaxies considered in \citet{lelli+14a}. It is prominent that the centrally star-bursting BCDs (black filled squares) and compact dwarf irregulars (blue filled squares) show systematically different locations compared to the dwarf irregulars, in which they have higher V$_{R_d}$/R$_d$ and smaller $\mu$$_0$ values. Given that centrally star-bursting BCDs are formed through major merging, this could imply that these galaxies are likely to have much steeper rotation curves owing to the active transportation of gas to the center of the merging remnants \citep{Watts&Bekki16}. On the other hand, BCDs with off-centered starbursts (black filled circles) are mostly located in the area containing dwarf irregulars (blue dots) on this plane; this could be explained based on the suggestion that off-centered star-bursting BCDs are formed from minor merging \citep{Watts&Bekki16}. The rotating dEs (red diamonds) in the Virgo cluster also populate a similar region of centrally star-bursting BCDs and compact dwarf irregulars, suggesting a possible connection between these different types of dwarf galaxies; i.e., compact dwarf irregulars and rotating dEs are suggested as descendants of BCDs \citep{lelli+14a}. EVCC 591 also has higher V$_{R_d}$/R$_d$ and smaller $\mu$$_0$ values compared to the dwarf irregulars and also occupies a similar region with centrally star-bursting BCDs and rotating dEs. This suggests that EVCC 591 has an evolutionary connection with BCDs and dEs in the sense that EVCC 591 may be a transitional galaxy on the way to becoming a dE evolving from a BCD in the cluster environment \citep[e.g.,][]{lisker06a}. \\

If dE(bc)s have evolved through quiescent phases with a low-level, central star formation after BCDs were formed by merging \citep[e.g.,][]{Bekki08}, they would have shallower rotation curves and less mass concentration owing to gas consumption. However, in their simulations, \citet{Watts&Bekki16} hinted that the descendants of BCDs exhibit little evolution on the plane of V$_{R_d}$/R$_d$ vs. $\mu$$_0$ compared to the BCDs \citep[see also][]{lelli+14a}. Consequently, we can suggest that the steep central rotation curve of EVCC 591 might be additional kinematical evidence supporting the merging event from which the central potential well is enhanced by efficient gas infall into the central region (\citealt{Bekki08, bekki15, Watts&Bekki16} and references therein). Such gas concentrations would also be responsible for the central star-formation activity of EVCC 591.\\

EVCC 591, locating near the core of the Virgo cluster, exhibits only a mild, central star-formation activity with H$\alpha$ and H$\beta$ emission lines (i.e., EW(H$\alpha$) $\sim$ 13.8  \AA\ and EW(H$\beta$)  $\sim$ 3.9 \AA\ from the SDSS archive; see Figure 1). Furthermore, HI gas was not detected in EVCC 591 from the Arecibo Legacy Fast ALFA (ALFALFA, \citealt{haynes18}) survey archive. The flux limit of ALFALFA is S$_{21}$ = 0.72 Jy km s$^{-1}$ \citep{haynes18}, which corresponds to the detection limit of the HI mass M$_{HI}$ = 10$^{7.7}$ M$_\odot$ at a distance of the Virgo cluster. Therefore, EVCC 591 is expected to have undergone ram-pressure stripping exerted by the intergalactic medium at the central region of the cluster. Ram-pressure stripping is more efficient in the outer part of the low-luminosity galaxy where gas is instantaneously removed. This prevents accretion of fresh gas into the galaxy center that triggers continuous, ongoing star formation events. In the case of dwarf galaxies in or near the cluster core, most of the gas is efficiently removed by ram-pressure stripping in a shorter time ($\sim$ 150 Myr) compared to the crossing time of the Virgo cluster \citep[$\sim$ 1.7 Gyr;][]{boselli&gavazzi06} owing to the shallow potential well of these galaxies. This causes a subsequent rapid stoppage of star-formation activity (\citealt{Boselli08a}). In this case, we anticipate that EVCC 591 will be rapidly transformed into a redder dE by passive evolution with a timescale of $\sim$ 1 Gyr after the central star formation ceases \citep[e.g.,][]{suh+10}. \\

\subsubsection{EVCC 516}

Unlike EVCC 591, EVCC 516 shows neither a prominent rotation in the velocity profile nor any direct kinematical feature, such as a KDC, signifying galaxy merging. In the current galaxy formation paradigm, the angular momentum originating from tidal torque and the quiescent cooling of gas within a virialized dark matter halo result in the formation of rotationally supported galaxies \citep{peebles69,WR78,Fall&Efstathiou80}. Thus, any physical mechanism for transforming rotationally supported systems into ones dominated by random motion is required. It has been established through numerical simulations that galaxy mergers can naturally account for the presence of pressure-dominated early-type galaxies \citep{Toomre77,Barnes88,Barnes92,Hernquist92,Cox06,Naab06,Jesseit09,Hopkins+13}. In particular, the mass ratio of the merger is the dominant factor in which an equal-mass merger is more favorable for the formation of slow- or non-rotating systems than an unequal-mass one \citep[e.g.,][]{Jesseit09}. Other secondary factors (e.g., merging geometry, gas fraction, and environment) also affect the final spin of the merger remnant \citep{Jesseit09,Choi&Yi17}. Note that a fraction of fast-rotating early-type galaxies can be transformed into slow-rotating systems on a considerably short timescale (i.e., less than 0.5 Gyr at z=0) from significant merger events \citep{Schulze+18}. In lower-mass regimes, non-rotating dwarf spheroidal galaxies also can be produced from mergers between disky dwarf galaxies (e.g., \citealt{Kazantzidis+11,Ebrova&Lokas15}; see also \citealt{Valcke+08}). Observational studies detected non-rotating BCDs inferring that mergers between gas-rich dwarf galaxies are plausible mechanisms for the formation of pressure-dominated BCDs (\citealt{Ostlin+99,Ostlin01,Blasco-Herrera+13} and references therein). In this respect, we are inclined toward speculating that the merging scenario also could explain the non-rotating feature of EVCC 516. \\

Numerical simulations demonstrated that environmental effects also dominantly influence galaxy rotation in a cluster environment \citep{Mastro+05,Choi&Yi17,Choi+18}. As ram-pressure stripping only directly affects the gas within a galaxy, stellar rotation of a galaxy would not be altered by this process alone. On the other hand, galaxy harassment, multiple gravitational interactions with massive member galaxies of a cluster, has a strong effect on the internal kinematics of the stellar population in the progenitor galaxy, decreasing its rotation \citep[e.g.,][]{moore98,Mastro+05,Smith+10}. If a galaxy entered the cluster at early epochs, it would go through the center of the cluster several times, and multiple events of harassment for several Gyrs lead to a large spin-down of the galaxy. For massive galaxies, slow rotators are mostly located in the central part of the cluster, but rarely exist in a low-density region of the cluster \citep{cappell11}. \citet{toloba15} also found that slow-rotating dEs are preferentially located in the inner region ($<$ 1 Mpc) of the Virgo cluster. They suggested that environmental effects of clusters such as harassment play a fatal role in reducing the rotation of dwarf galaxies. Furthermore, after several passes through the cluster center, the ram-pressure stripping effect also mostly removes any gas even in the galaxy center, resulting in the formation of a red dE galaxy \citep{Boselli08a,boselli08b}. \\

However, this would not be relevant to the case of EVCC 516. As EVCC 516 is located far outside the X-ray halo of the Virgo cluster (see Figure 1), it is highly impossible that all the gas of this galaxy can be immediately removed by the effects of the ICM. This is supported by the strong ongoing star formation activity of EVCC 516 at its central region (i.e., EW(H$\alpha$) $\sim$ 101.5 \AA\ and EW(H$\beta$) $\sim$ 27.4 \AA\ from the SDSS archive) and a large amount of HI gas (i.e., HI mass $\sim$ 10$^{7.8}$ M$_{\odot}$ and HI gas mass fraction = HI mass/(stellar mass + HI mass) $\sim$ 0.08 from the ALFALFA survey archive). Therefore, we suggest that EVCC 516 has only recently fallen into the outskirts of the Virgo cluster and has not passed through the center of the Virgo cluster; therefore, environmental effects are not the dominant drivers of the internal kinematics of EVCC 516. Moreover, we could not find any potential neighbor galaxy within a projected distance of 300 kpc from EVCC 516 considering the galaxies in the EVCC \citep{kim14}. In this case, EVCC 516 could be rather considered as an analog of a dE(bc) found in an isolated environment (e.g., IC 225, \citealt{gu06}).\\

As mergers are often associated with intense starburst episodes, one may expect that this process quickly exhausts the gas reservoir of a merger remnant. However, according to the numerical simulations of \citet{Bekki08}, BCDs formed from mergers between gas-rich dwarfs were surrounded by massive extended HI gas envelopes, indicating that a substantial amount of HI gas remains bound. Furthermore, the HI envelope around the BCDs in an isolated environment can be retained over a long timescale ($>$ 10 Gyr) as stellar feedback in BCDs is not strong enough to expel all the gas out of the potential well of the BCDs (\citealt{Ferrara+00,Tajiri&Kamaya02,SanchezAlmeida+08,lelli+14a,lelli+14b,Valcke+08} for spheroidal dwarf galaxies). This is supported by the fact that the overall gas-depletion timescales of BCDs are much longer than the starburst timescales \citep[e.g., a few 100 Myr,][]{McQuinn+10}, implying that BCDs do not consume their entire gas reservoirs \citep{lelli+14a}. Furthermore, isolated dwarf galaxies do not quench by secular processes unless they are close to a massive galaxy \citep{Geha+12}.\\

It has been suggested that isolated star-bursting dwarf galaxies without rotational features exhibit episodic star-formation histories \citep{stinson+07,Valcke+08,Revaz+09,Schroyen+11,Schroyen+13}. In this case, the isolated BCDs during quiescent phases can be observed as objects without major starburst characteristics, referred to as quiescent BCDs \citep[QBCD;][]{SanchezAlmeida+08,SanchezAlmeida+09}. QBCDs might be observed as transitional dwarf galaxies in between BCDs and dEs \citep{SanchezAlmeida+08}. Many fractions of QBCDs exhibit morphologies hinting centrally concentrated star formation activity, which are akin to those of dE(bc)s, in contrast to BCDs showing global, intense star-bursting features \citep[see Figure 3 of][]{SanchezAlmeida+08}. It has been also suggested that a repetitive sequence of BCD and QBCD phases is maintained during several Hubble times, which is a consequence of the replenishment of fresh gas by inflows from the HI gas envelope and subsequent outflows owing to starbursts (see \citealt{SanchezAlmeida+08,SanchezAlmeida+09,SanchezAlmeida+14} for details; see also \citealt{Valcke+08,lelli+14a,lelli+14b}). Using the selection criteria proposed by \citet{SanchezAlmeida+08}, we confirmed that EVCC 516 has the general characteristics of the QBCD population, i.e., $g$-$r$ = 0.48 mag from the EVCC; 12+log (O/H) = 8.42, log ([N II]$\lambda$6584/H$\alpha$) = -0.95, and log ([O III]$\lambda$5007/H$\beta$) = 0.5 calculated from the SDSS spectroscopic data for EVCC 516.\\

If gas inflows from the HI gas envelope occur, the gas kinematics can become disturbed \citet{Verbeke+14}. In particular, the kinematics of infalling gas can exhibit rotation at the galaxy center (see \citealt{Verbeke+14} for details). It is interesting to note that this characteristic predicted in numerical simulations considering gas accretion appears to be very similar to the kinematical feature shown in EVCC 516, i.e., a possible rotation feature of gas components at the galaxy center (see Figure 4g). We also examined the shape of the HI spectrum of EVCC 516 extracted from the ALFALFA survey archive. However, we only found a single peak, narrow HI profile, which suggests no significant rotation of HI gas surrounding EVCC 516. \\

It is worth noting that external gas infall was suggested based on the HI observations (see \citealt{Verbeke+14} and references therein). For example, NGC 5253, a nearby BCD \citep{Karachentsev+07}, is surrounded by a large-scale ($\sim$ 2 to 3 times its optical size), massive ($\sim$ 10$^8$ M$_{\odot}$) HI structure \citep{Lopez-Sanchez+12}. \citet{Lopez-Sanchez+12} concluded that NGC 5253 is probably experiencing the infall of a diffuse, low-metallicity HI cloud, which triggered the powerful central star-burst activity in this galaxy. Surprisingly, \citet{DeRijcke+13} also found a massive ($\sim$ 10$^7$ M$_{\odot}$) HI gas cloud surrounding FCC 046, a possible dE(bc) with signatures of centrally concentrated recent star-formation activities  \citep{drinkwater+01,DeRijcke+03}, located far outside the X-ray halo of the Fornax cluster. Wide-field, deep HI mapping around EVCC 516 will be necessary for thorough understanding of the gas distribution and evolution of this galaxy; this is a possible example of a QBCD having a large reservoir of diffuse, neutral gas surrounding the galaxy.  \\

There is a technical and observational caveat for studying internal kinematics in the long-slit spectra \citep{toloba14a}. If the long-slit is not aligned with respect to the rotation field of the galaxy, long-slit spectroscopy would have missed rotational features or kinematic anomalies signifying merging events. In fact, slow-rotating elliptical galaxies formed from equal-mass mergers tend to exhibit significant misalignments between photometric and kinematic major axes, while fast-rotating systems do not show such kinematical misalignments \citep{Cappellari+07,Emsellem+07}. Moreover, the minor-axis rotation of a dwarf spheroidal galaxy can also originate from the merger between disk dwarf galaxies (\citealt{Ebrova&Lokas15}; see also \citealt{Schulze+18} for prolate rotation of early-type galaxies). However, owing to the limitation of our long-slit observations on the photometric major axis, we might not have captured the internal kinematics of the minor axis or the whole galaxy. This caveat can be better alleviated with integral-field spectroscopic observations, but observations of large samples of dE(bc)s using this technique are not reported yet. \\

\subsection{Evolutionary Connections with BCDs and dEs}

It has been noted that there are some structural similarities between star-bursting BCDs and dEs  \citep[e.g.,][]{kormendy85,meyer14}, while the evolutionary connection between these two-types of galaxies is still under debate. In particular, large fractions of BCDs show morphologies with a smooth elliptical appearance and centrally concentrated star-bursting region \citep[e.g., nuclear elliptical (nE) BCD, ][]{loose&thuan86}, which are fairly similar to the typical morphology of dE(bc)s. Therefore, it would be possible to consider that a fraction of dE(bc)s in the cluster environment are transitional objects possibly evolved from BCDs and are on their way to becoming dEs. In this regard, it is worth comparing the structural properties of BCDs with those of dEs and dE(bc)s. \citet{meyer14} showed that low surface brightness (LSB) components of BCDs in the Virgo cluster, which are dominated by underlying old stellar population, structurally resemble the compact Virgo dEs.\\

The structural parameters of EVCC 516 and EVCC 591 were compared with those of other types of dwarf galaxies. As comparison samples, we used the dEs, dE(bc)s, LSB components of BCDs, and BCD candidates in the Virgo cluster extracted from Figures 6 and 7 of \citet{meyer14} using DEXTER (see \citealt{meyer14} for details of the sample galaxies). Figure 6 illustrates the scaling relations of the absolute magnitude vs. R$_{eff,r}$ (left panel) and $<$$\mu$$>$$_{eff,r}$ (right panel) for the galaxies. As shown in the left panel of Figure 6, dE(bc)s (green asterisks) including EVCC 591 and EVCC 516 are overlapped with the BCD candidates (blue filled squares) and reside on the linear extension of the distribution of LSB components of BCDs (red filled squares), forming a continuous sequence. At a given magnitude, all of these galaxies are systematically compact (i.e., small R$_{eff,r}$) compared to the majority of dEs (black dots). This indicates that BCDs and dE(bc)s have high surface brightness compared to dEs with similar magnitudes (see right panel of Figure 6). As noted by \citet{meyer14}, the Virgo BCDs could evolve into more compact dE(bc)s and dEs than those currently observed in the Virgo cluster once their starbursts are halted by environmental effects.\\

We also measured the structural parameters (R$_{eff,r}$ and $<$$\mu$$>$$_{eff,r}$) of seven dEs with KDCs (VCC 510, VCC 917, VCC 1183, VCC 1261, VCC 1453, VCC 1475, and VCC 1545) compiled from literature (\citealt{thomas06,Chilingarian+07,Chilingarian09,koleva+11,toloba14a,Guerou15}; see Figure 1 for their spatial distribution) using the SDSS $r$-band images. Interestingly, as shown in Figure 6, dEs with KDCs (purple filled circles), which are considered to be formed from galaxy merging, also appear to be offset from the average distribution of dEs. Moreover, these galaxies lie within the distribution of dE(bc)s, being relatively more compact compared to the majority of dEs with similar magnitudes\footnote{We found that one dE with KDC, VCC 1475, has very small R$_{eff,r}$ and high surface brightness and is located outside of the mean distribution of galaxies in Figure 6. This is the result of its relatively large and bright core in its visual appearance in the SDSS color image, while other dEs with KDCs contain only a small nucleus at the center.}. This suggests that the dEs with KDCs may be structurally related to the dE(bc)s. \\

Bekki (2008) predicted a significant positive age gradient and negative metallicity gradient of the newly formed stellar population in a merger remnant. Therefore, regarding galaxy merging as a formation channel of dE(bc)s and their possible evolutionary link with dEs, it would be interesting to examine if dEs with KDCs also show similar age and metallicity gradients. Information on the age and metallicity gradient of all the seven dEs with KDCs in the Virgo cluster is available in literature. The KDCs in six dEs (VCC 917, VCC 1183, VCC 1261, VCC 1453, VCC 1475, and VCC 1545) contain younger and possibly more metal-rich populations than the main body of the galaxy \citep{Chilingarian+07,Chilingarian09,koleva+11,toloba14a}. The KDC of VCC 510 also hints at the existence of a younger stellar population but with slightly lower metallicity, compared to the outer part of the galaxy \citep{thomas06}. It has been suggested that the KDC consisting of a young, metal-rich stellar population can be formed through a dissipative merger between gas-rich galaxies \citep{Chilingarian+07}. It is likely, therefore, that dE(bc)s with KDCs are evolutionarily connected with normal dEs having KDCs in the cluster environment. If external processes remove the gas from the dE(bc)s with KDCs in the cluster, the galaxies are subsequently quenched to dEs, but their overall stellar kinematic features, including KDCs, can be maintained over a long timescale  \citep[e.g., a few Gyr,][]{tsatsi15}. \\

\section{SUMMARY AND CONCLUSIONS}

In this paper, we have presented the internal kinematics of two dE(bc)s in the Virgo cluster, EVCC 591 and EVCC 516, based on Gemini GMOS long-slit spectroscopic observations. The two galaxies are located in different environments within the cluster; EVCC 591 and EVCC 516 are at the cluster core and outskirts of the cluster, respectively. We obtained velocity profiles along the major axis out to, at least, $\sim$ 1 R$_{eff}$. The velocity profile of EVCC 591 indicates an off-centered nucleus decoupled from the overall distinct rotation of the main body of the galaxy. In addition to the presence of the KDC, EVCC 591 also exhibits a steep central rotation curve compared to the rest of the galaxy, implying a strong central concentration of the dynamical mass. On the other hand, EVCC 516 shows no prominent rotation in its velocity profile, indicating a pressure-dominated system. Considering the observed kinematical properties and central star formation, we are inclined toward concluding that the formation of our two dE(bc)s is somehow connected to the merging event between dwarf galaxies with available gas reservoirs. \\

The dEs with KDCs in the Virgo cluster bear a striking resemblance to dE(bc)s in general with regard to their structural parameters. Furthermore, most dEs with KDCs show hints of a positive age gradient and negative metallicity gradient, which are consistent with the properties of BCDs and dE(bc)s. Therefore, it is almost certain that dEs with KDCs are descendants of dE(bc)s formed by merging after being quenched in the cluster environment. The internal kinematics of the large samples of dE(bc)s will further provide constraints on possible evolutionary connections between dE(bc)s and dEs with KDC. \\

If one were to assume that galaxy merging is a formation channel for at least some of the dE(bc)s found in the Virgo cluster, this event would have occurred in low-density environments such as galaxy group and field before they subsequently fell into the cluster environment. In contrast to the cluster environment, mergers and interactions between galaxies are more frequent in low-density environments with relatively low velocity dispersions \citep{Binney&Tremaine+87, boselli&gavazzi06}. The recent star formation history of dE(bc) may be different owing to the amount of remaining gas at the center of the galaxy when a galaxy falls into the cluster \citep{Shaya&Tully84,lisker06a,Tully&Trentham08,DeRijcke+13}. The obvious difference in the observed properties (e.g., location in the cluster, strength of star-formation activity, and HI gas detection) between EVCC 591 and EVCC 516 may be largely related to the external influences of the cluster environment on the galaxy.\\

It is interesting to note that dE(bc)s are abundantly found outside of clusters. In fact, galaxy groups show a much larger fraction of dE(bc)s (i.e., $\sim$ 70$\%$ of dEs; \citealt{Tully&Trentham08,pak+14}) than the Virgo cluster ($\sim$ 5$\%$ of all dEs; \citealt{lisker06a,lisker07,meyer14}). Moreover, the most recent study based on the extensive sample at z $<$ 0.01 also revealed that the population of dE(bc)s is also ubiquitous in the field environment, and its fraction is much higher ($\sim$ 90$\%$; S.-C. Rey et al. in prep.) than that in cluster and group environments. This provides observational evidence that low-density environments are very favorable to the formation of dE(bc)s. In this respect, studies in the internal kinematics of dE(bc)s in low-density environments are highly expected. Coupled with detailed internal kinematics of large samples of dE(bc)s in the cluster, this approach has the potential to place strong constraints on the nature of dE(bc)s and their evolution. \\

\clearpage

\acknowledgments{We are grateful to the anonymous referee for helpful comments and suggestions that improved the clarity and quality of this paper. We thank Martha Haynes for providing the ALFALFA HI spectrum data used in this study. S. C. R., acting as the corresponding author, acknowledges support from the Basic Science Research Program through the National Research Foundation of Korea (NRF) funded by the Ministry of Education (2018R1A2B2006445). Support for this work was also provided by the NRF to the Center for Galaxy Evolution Research (2017R1A5A1070354). This work was supported by K-GMT Science Program (PID: [GN-2015A-Q-202]) of Korea Astronomy and Space Science Institute (KASI). Based on observations obtained at the Gemini Observatory processed using the Gemini IRAF package, which is operated by the Association of Universities for Research in Astronomy, Inc., under a cooperative agreement with the NSF on behalf of the Gemini partnership: the National Science Foundation (United States), the National Research Council (Canada), CONICYT (Chile), Ministerio de Ciencia, Tecnolog\'{i}a e Innovaci\'{o}n Productiva (Argentina), Minist\'{e}rio da Ci\^{e}ncia, Tecnologia e Inova\c{c}\~{a}o (Brazil), and the Korea Astronomy and Space Science Institute (Korea).} \\

\clearpage

\begin{figure*}
\plotone{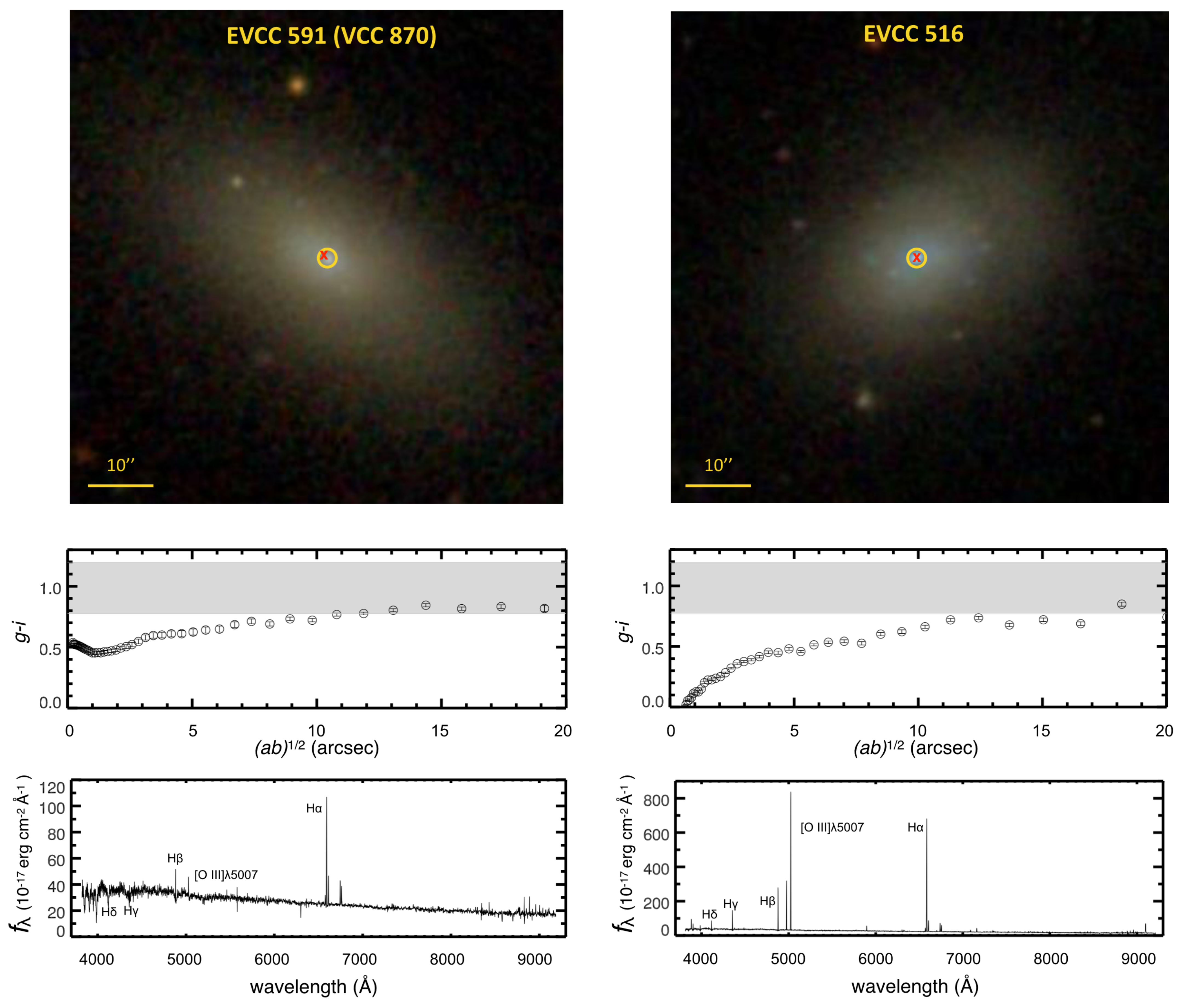} 

\caption{Two target dE(bc)s, EVCC 591 (left) and EVCC 516 (right), for long-slit spectroscopic observations. (Top) SDSS $g$, $r$, and $i$ composite color images of galaxies. North is up and east is to the left. The central small circle denotes the fiber aperture with 3$\arcsec$ diameter of SDSS spectroscopic observations. The red cross is the photometric center of the galaxy obtained from the EVCC. (Middle) Radial $g$-$i$ color profiles of galaxies obtained from SDSS images. The radius is calculated from semimajor $\it{(a)}$ and semiminor $\it{(b)}$ axes as ($ab$)$^{1/2}$. The error bars denote the uncertainties calculated from magnitude errors. The gray-shaded areas represent the 2$\sigma$ range of the $g$-$i$ colors of normal dEs without blue cores at their respective magnitude derived from the color-magnitude relation of the Virgo cluster using the EVCC \citep{kim14}. (Bottom) SDSS optical spectra of galaxies at their centers within a 3$\arcsec$ fiber diameter, displaying Balmer and [O III]$\lambda$5007 emission lines.}

\end{figure*} 

\clearpage

\begin{figure*}
\plotone{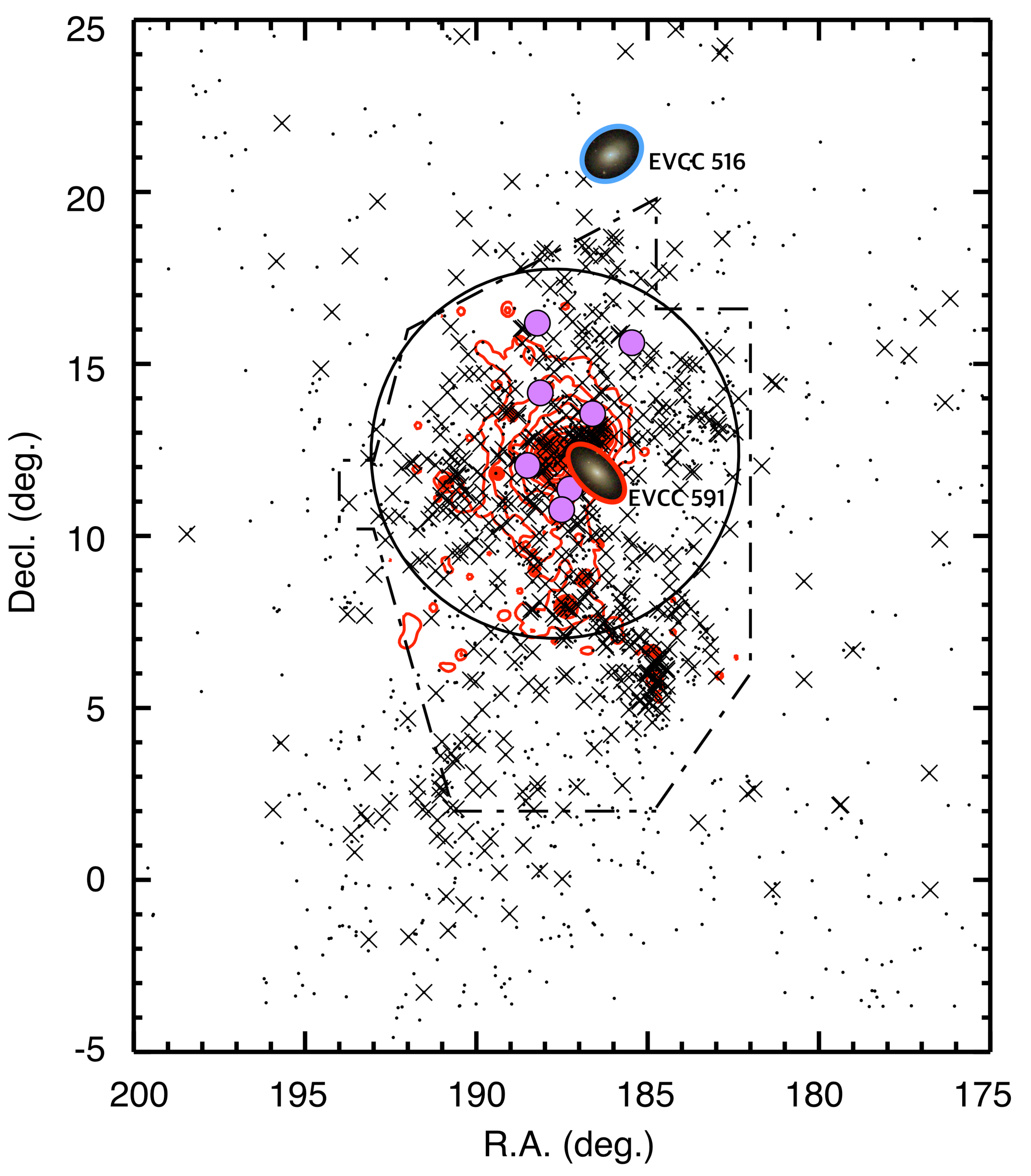} 

\caption{Projected spatial distribution of galaxies in the EVCC. The positions of two target dE(bc)s, EVCC 591 and EVCC 516, in long-slit spectroscopic observations are marked using SDSS images. The crosses denote the dEs, whereas dots are other types of galaxies included in the EVCC. Seven purple filled circles are dEs with KDCs reported in literature (see Sec. 3.3 for details). Red contours represent the X-ray diffuse emission distribution of the cluster from the ROSAT. The large circle indicates the virial radius of Virgo A (5$\degr$.37 = 1.55 Mpc) defined by \citet{Ferra12}, considering a distance of 16.5 Mpc to the Virgo cluster from us \citep{Jerjen+04,mei07}. The region of the Virgo Cluster Catalog (VCC; \citealt{binggeli85}) is represented by the dot-dashed contour.}

\end{figure*} 

\clearpage


\begin{figure*}
\epsscale{1.0}
\plotone{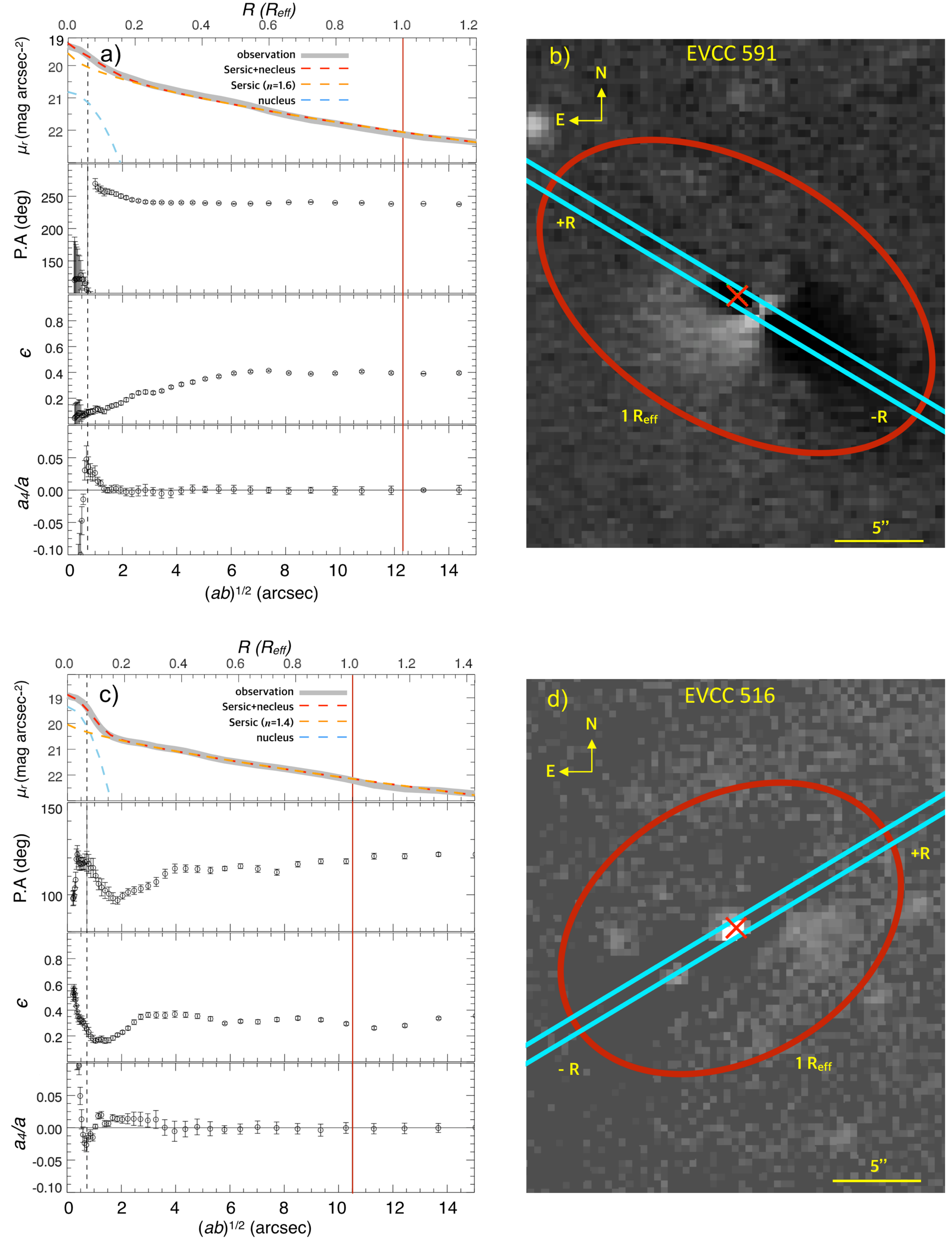} 

\caption{(a) and (c): Profiles of isophotal parameters from ellipse fitting of  EVCC 591 and EVCC 516. The radius is calculated from semimajor $(a)$ and semiminor $(b)$ axes as ($ab$)$^{1/2}$. From top to bottom, profiles of surface brightness, position angle, ellipticity, and $a$$_4$/$a$ parameter are presented. In the surface brightness profile, the best-fitted model line is represented by the red dashed line, which is a combination of the Gauss function for the nucleus (blue dashed line) and S\'ersic function for the main body (orange dashed line). The red vertical line indicates the effective radius of a galaxy. The error bars are uncertainties measured by IRAF $\it{ELLIPSE}$ fitting. The dotted vertical line in each panel indicates a median seeing radius ($\sim$ 0$\arcsec$.7) of the SDSS $r$-band image within which data points are excluded from ellipse fitting. (b) and (d): SDSS residual images of EVCC 591 and EVCC 516 obtained by subtracting two-dimensional smooth model images constructed by ellipse fitting from the original SDSS $r$-band images. The red cross is the photometric center of the galaxy obtained from the EVCC. The red ellipse corresponds to an elliptical isophote with an effective radius in the semimajor axis. Cyan lines indicate the width (1$\arcsec$) and orientation of the long slit used in the spectroscopic observations. The positive and negative radial distances from the galaxy center are denoted by +R and -R, respectively.}

\end{figure*} 

\clearpage

\begin{figure*}
\epsscale{1.2}
\plotone{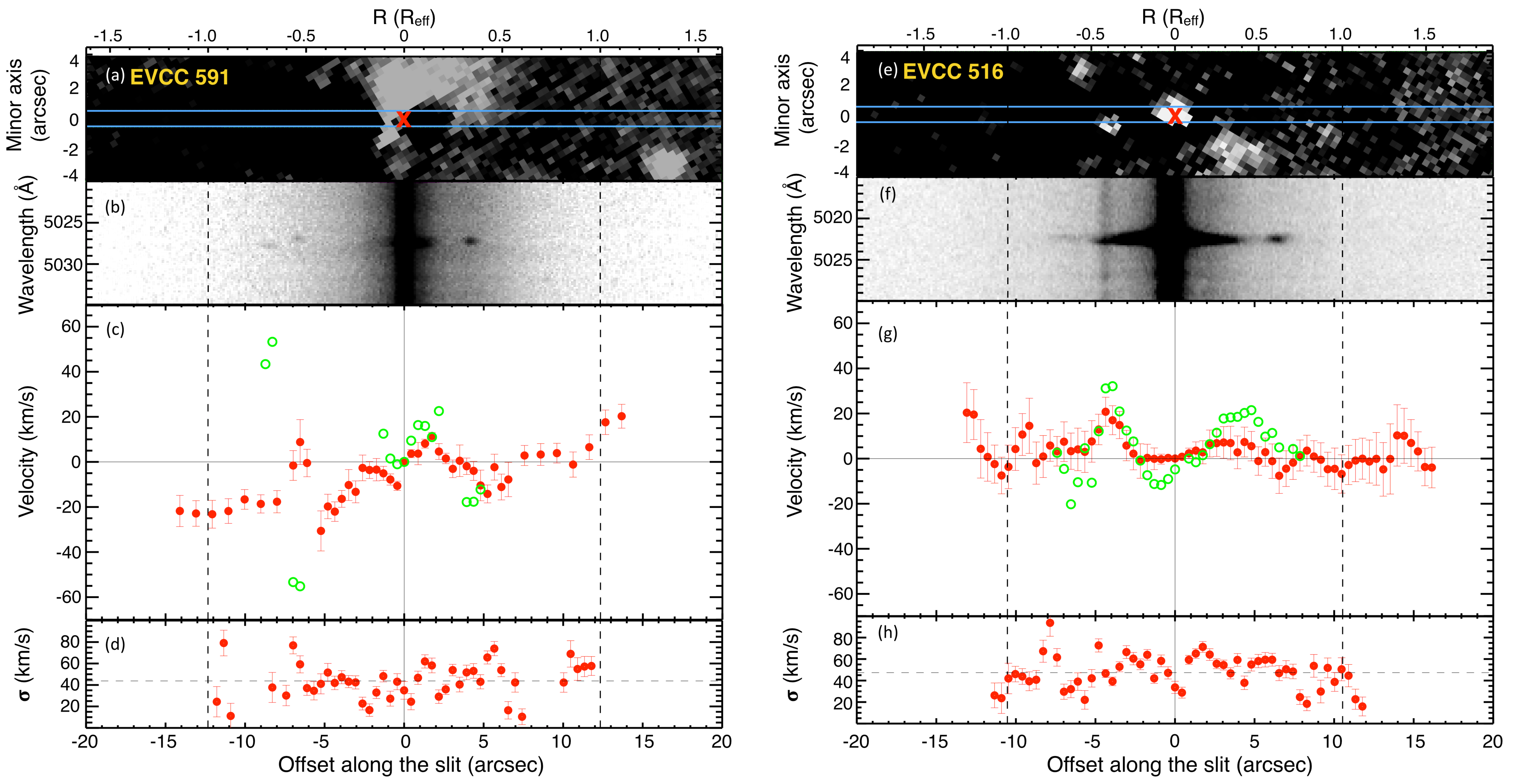} 

\caption{Observable profiles of EVCC 591 (left panels) and EVCC 516 (right panels) as a function of distance from the galaxy center along the major axis. In all panels, dashed vertical lines indicate the effective radius of galaxies. (a) and (e): SDSS $r$-band residual images obtained by subtracting two-dimensional smooth model images constructed by ellipse fitting. The blue lines are the footprints of the long slit of our spectroscopic observations. The red cross is the photometric center of the galaxy obtained from the EVCC. (b) and (f): Two-dimensional spectrum around [O III]$\lambda$5007 emission line. (c) and (g): Velocity profiles of stellar (red circles) and ionized gas (green circles) components with respect to the systemic velocity of galaxies. Error bars represent the 1$\sigma$ uncertainties returned from the $\it{FXCOR}$ task. (d) and (h): Stellar velocity dispersion profiles. The dashed horizontal line denotes the mean values (44.01 km s$^{-1}$ and 47.38 km s$^{-1}$ for EVCC 591 and EVCC 516, respectively) of velocity dispersions within the effective radius. Error bars represent the 1$\sigma$ uncertainties returned from the $\it{FXCOR}$ task.}

\end{figure*}

\begin{figure*}
\plotone{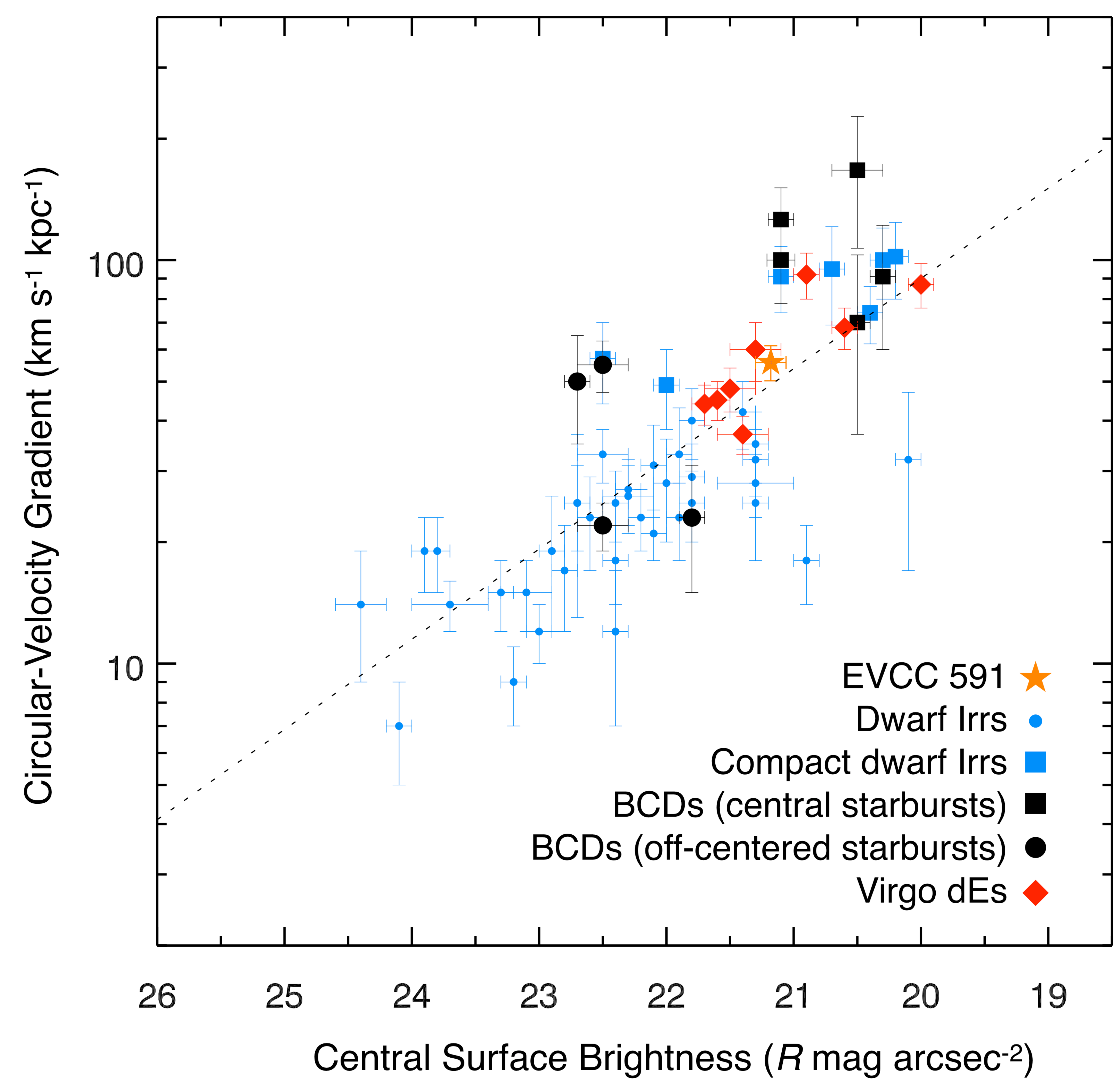} 

\caption{Circular-velocity gradient (V$_{R_d}$/R$_d$) versus inclination-corrected $R$-band central surface brightness ($\mu$$_0$) of various types of dwarf galaxies extracted from \citet{lelli+14a}. Black filled squares and circles are BCDs with a central starburst and off-centered starburst, respectively. Blue dots and blue filled squares are typical dwarf irregulars and compact dwarf irregulars, respectively. Red diamonds are rotating dEs in the Virgo cluster. EVCC 591 is also marked by a orange star. The dashed line is a linear fit to the data obtained from \citet{lelli+14a}.}

\end{figure*} 

\clearpage

\begin{figure*}
\plotone{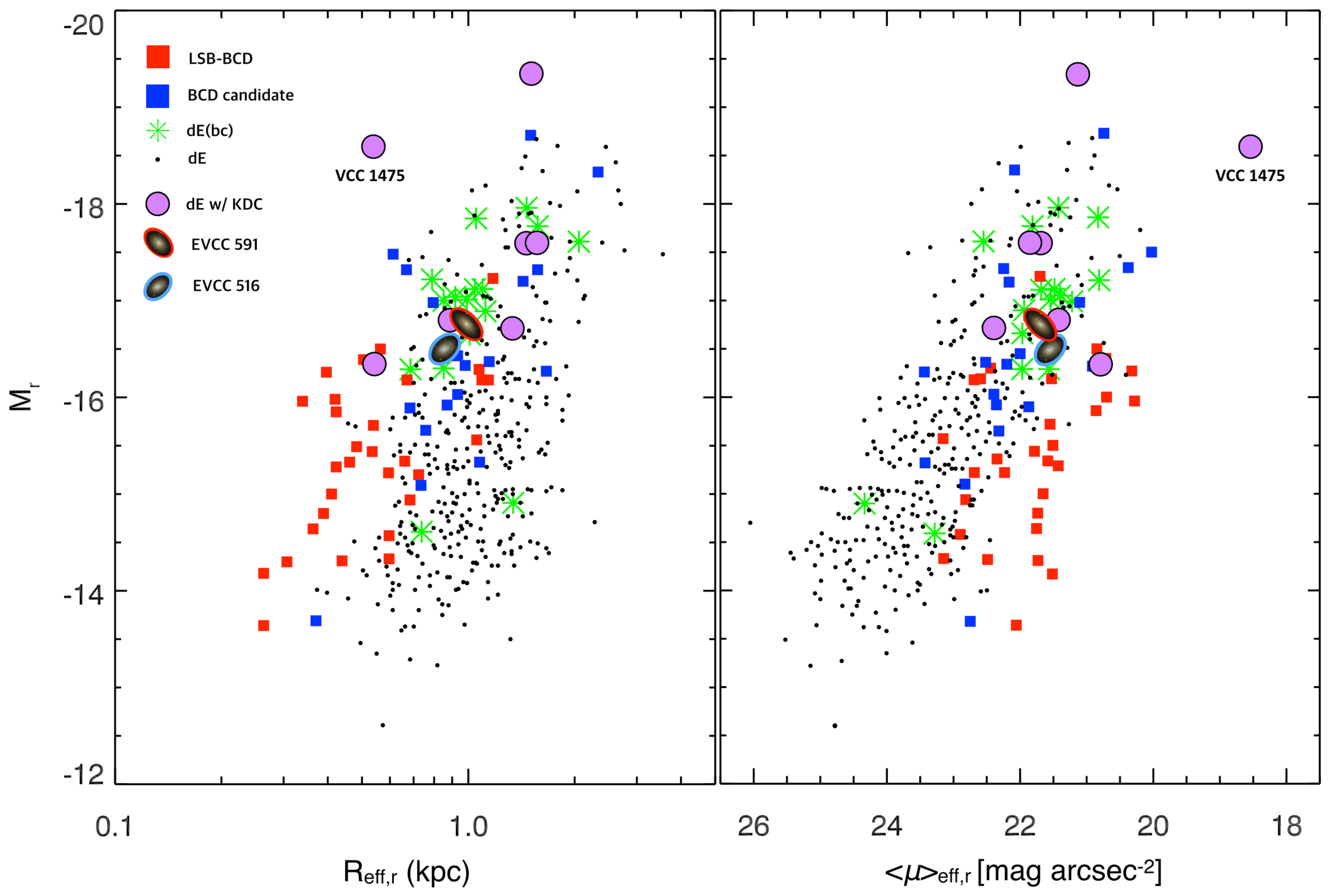} 

\caption{Distribution of dwarf galaxies of various types in scaling relations: absolute $r$-band magnitude (M$_r$) vs. effective radius (R$_{eff,r}$) (left) and absolute $r$-band magnitude (M$_r$) vs. mean effective surface brightness in $r$-band ($<$$\mu$$>$$_{eff,r}$) (right). EVCC 591 and EVCC 516 are denoted by SDSS images. As a comparison sample, we plot the dEs (black dots), dE(bc)s (green asterisks), LSB components of BCDs (red filled squares), and BCD candidates (blue filled squares) in the Virgo cluster extracted from \citet{meyer14}. Seven dEs with KDCs in the Virgo cluster compiled from literature are also marked with purple filled circles.}

\end{figure*} 

\clearpage

\begin{deluxetable}{cccccccccccc} 
\tablewidth{0pt}
\tablecaption{Basic Characteristics of the Target Galaxies}
\tablehead{
\colhead{Object}           & \colhead{VCC}      & 
\colhead{R.A. (deg)}          &
\colhead{Dec. (deg)}          & \colhead{$cz$ (km/s)} &
\colhead{$r$}    & \colhead{$i$}  & \colhead{$g-i$} & \colhead{$M$$_\star$/$M$$_\odot$}   & \colhead{$R$$_{eff,r}$ ($\arcsec$, kpc)}  & \colhead{$<$$\mu$$>$$_{eff,r}$} & \colhead{$\epsilon$$_{eff,r}$}
\\
\colhead{(1)}           & \colhead{(2)}      &  \colhead{(3)}          & \colhead{(4)} & \colhead{(5)}  &
\colhead{(6)}            & \colhead{(7)} & \colhead{(8)} & \colhead{(9)} & \colhead{(10)} & \colhead{(11)} & \colhead{(12)} }

\startdata
EVCC 591 & VCC 870   & 186.5224 & 11.8123 & 1194 & 14.25 & 13.87 & 0.92 & 10$^{9.1}$ & 12.3$\pm$0.1, 0.98$\pm$0.01 & 21.70$\pm$0.13 & 0.262$\pm$0.003\\
EVCC 516 & {\nodata} & 186.0920 & 21.1602 & 955 & 14.48 & 14.20 & 0.76 & 10$^{8.8}$ & 10.5$\pm$0.1, 0.84$\pm$0.01 & 21.59$\pm$0.12 & 0.274$\pm$0.004\\
\enddata
\tablecomments{(1) Galaxy name in the EVCC; (2) Galaxy name in the VCC; (3) and (4) Right ascension (J2000) and declination (J2000) from the EVCC, respectively; (5) Radial velocity from the SDSS; (6) SDSS $r$-band magnitude from the EVCC; (7) SDSS $i$-band magnitude from the EVCC; (8) SDSS $g-i$ color from the EVCC; (9) Stellar mass; (10) Effective radius (in arcsec and kpc) in the SDSS $r$-band image; (11) SDSS $r$-band mean effective surface brightness within the effective radius; (12) SDSS $r$-band uncertainty-weighted mean ellipticity within the effective radius} 

\end{deluxetable}

\begin{deluxetable}{lcccccccc} 
\tablewidth{0pt}
\tablecaption{Instrumental Configuration of the Observations}
\tablehead{
\colhead{	}           & \colhead{EVCC 591}      & 
\colhead{	}          &
\colhead{EVCC 516}         
\\
\colhead{ }           & \colhead{(1)}      &  \colhead{	}          & \colhead{(2)} }

\startdata
Grating (line/mm) & {	} & B1200 & {  }\\
Wavelength range (\AA) & {    } & 4680 - 6150 & {  } \\
Slit width ($\arcsec$) & {    } & 1 & {  } \\
Exposure time (sec) & {	} & 9 $\times$ 1200 & {} \\
P.A.$^1$ (deg) & 239 & {  } & 121 \\
(S/N)$_o$$^2$ & 15 & { } & 20 \\
(S/N)$_{eff}$$^3$ & 7 & {  }& 5
\enddata
\tablecomments{1. Position angle, measured north-east, for the placement of the long slit; 2. S/N in the bin at galaxy center; 3. S/N in the bin at effective radius}

\end{deluxetable}
\clearpage


\end{document}